\documentclass[12pt]{article}

\usepackage{graphicx}
\usepackage{epsfig}
\usepackage{amsfonts}
\usepackage{amssymb}

\def\oneone{\rlap 1\mkern4mu{\rm l}}
\def\ep{{\epsilon}}

\begin{document}

\title{
Asymptotically flat, stable  black hole
solutions in \\Einstein--Yang-Mills--Chern-Simons theory}
 \vspace{1.5truecm}
\author{{\large Yves Brihaye,}$^{\dagger}$
{\large Eugen Radu}$^{\star \diamond }$
and {\large D. H. Tchrakian}$^{\star \diamond }$ 
\vspace*{0.2cm}
\\
$^{\dagger}${\small Physique-Math\'ematique, Universit\'e de Mons, Mons, Belgium} 
   \\
$^{\star}${\small  Department of Computer Science,
National University of Ireland Maynooth,
Maynooth,
Ireland} \\
$^{\diamond}${\small School of Theoretical Physics -- DIAS, 10 Burlington
Road, Dublin 4, Ireland }} 
\date{\today}
\maketitle
\begin{abstract}
We construct finite mass, asymptotically flat  black hole
solutions in  $d=5$ Einstein--Yang-Mills--Chern-Simons theory. 
Our results indicate
the existence of a second order phase transition
between Reissner-Nordstr\"om solutions and the non-Abelian black holes which  generically  are 
thermodynamically preferred.
Some of the non-Abelian configurations are also stable 
under  linear, spherically
symmetric perturbations.
 In addition
a solution in closed form describing an extremal black hole with non-Abelian hair is found for a special
value of the Chern-Simons coupling constant.
\end{abstract}

\noindent{\textbf{~~~Introduction.--~}}
The so called 
"no-hair" conjecture \cite{wheeler} states that an asymptotically flat,
stationary black hole is uniquely
described in terms of a small set of asymptotically measurable quantities. 
However, in recent years counterexamples to this conjecture 
were found in several theories, most of them containing non-Abelian matter fields. 
The first non-Abelian "hairy" black hole solutions within the framework of $d=4$ SU(2)  
Einstein-Yang-Mills (EYM) theory, were presented in \cite{89}.
Although these solutions were static and spherically symmetric with  vanishing Yang-Mills (YM) 
charges, they were different from the Schwarzschild black hole and,
therefore, not characterized exclusively by their total mass.  
However,   all known asymptotically flat
EYM solutions are perturbatively unstable and thus they do not contradict
the spirit of the  "no-hair" conjecture (see the review \cite{Volkov:1998cc}).
When considering instead a number $d>4$ of spacetime dimensions,
no finite energy asymptotically flat non-Abelian solutions are found 
\cite{Volkov:2001tb,Okuyama:2002mh}
unless the action is supplemented with string-theory inspired higher order  YM 
curvature terms \cite{Brihaye:2002hr}, 
in which case,  again the solutions are classically unstable.

In odd spacetime dimensions, the usual gauge field action can be augmented instead 
by a Chern-Simons (CS) term. Such (CS) terms appear
in various supersymmetric theories.  In the Abelian case this  
leads to some new features for rotating black holes only~\cite{Gauntlett:1998fz}.
 In the non-Abelian case however,   
the CS term  can  affect the properties of solutions even in the static, spherically symmetric case. 
For the $d=5$ case considered here, this allows the construction of finite mass, asymptotically flat, non-Abelian  
black hole solutions which generically turn out to be thermodynamically favoured over the
Reissner-Nordstr\"om Abelian configurations. Moreover, some of these solutions are stable against linear, spherically
symmetric perturbations.
 
\noindent{\textbf{~~~The model.--~}}In a $4+1$ dimensional spacetime, the smallest simple gauge group 
supporting a nonvanishing CS term is $SO(6)$. 
Then we consider a general EYMCS
theory with Lagrangian
  \begin{eqnarray}
\label{lagr}
 &&{\cal L}=
 \frac{1}{16 \pi G}R\, {*\oneone} - \frac{1}{4}{*F}\wedge F 
 -\kappa  \ep_{I_1\cdots I_6} 
\Big(F^{I_1 I_2} \wedge  F^{I_3 I_4}\wedge  A^{I_5 I_6} 
\\
\nonumber
  && {~~~~~~~~~~~~~~~~~~~~~}
  -
g F^{I_1 I_2}\wedge  A^{I_3 I_4}\wedge  A^{I_5 J} \wedge A^{J I_6}
 +\frac{2}{5} g^2  A^{I_1 I_2}\wedge  A^{I_3 J}\wedge  A^{J I_4}\wedge 
A^{I_5 K} \wedge A^{K I_6} 
\Big),  \quad
\end{eqnarray}
where $A^{IJ}$ are the $SO(6)$ gauge fields,  $F^{IJ}=dA^{IJ}+g A^{IK}\wedge A^{KJ}$, 
$G$ is gravitational constant, $\kappa$  the CS coefficient and 
 $g$ the gauge coupling constant. 

Our solutions are spherically symmetric with  a line element
\begin{eqnarray}
\label{metric-gen}  
ds^{2}=-  f_0 dt^{2}+f_1 dr^2 +f_2 d\Omega^2_{3}, 
\end{eqnarray} 
where $f_i$ are functions of  $ r$ and $t$ in general and
$d\Omega^2_{3}=d\psi^2+ \sin^2\psi (d\theta^2+\sin^2 \theta d\phi^2)$ is the line element
of the three dimensional sphere. 

The general spherically symmetric Ansatz for the $SO(6)$ YM field is given \cite{Brihaye:2009cc}.   In the present
work we restrict ourselves to a consistent truncation, $SO(4)\times SO(2)$, of that Ansatz. Apart from simplifying
the picture, prominent new physical features are conveniently exposed in this truncation  of the Ansatz,
stated as, 
\begin{eqnarray}
\label{YMansatz}
A=\frac{1}{g}
\left(
\frac{w(r,t)+1}{r} \Sigma_{ij}\frac{x^i}{r}dx^j
+V(r,t)\Sigma_{56}dt
\right),~~{\rm with~~} i,j=1, \dots,4\,,
\end{eqnarray}
$\Sigma_{ij}$ being the representation matrices of $SO(4)$, and $\Sigma_{56}$ of the $SO(2)$, subalgebras
of $SO(6)$.  The Cartesian coordinates $x^i$ are related to the spherical coordinates $(r, \psi, \theta, \phi)$
as in flat space. 
 
Starting with static solutions, we have found it convenient to choose 
for the metric function in (\ref{metric-gen}),
$ f_1=1/N(r),~f_2=r^2,~f_0=N(r)\sigma^2(r)$, 
with $N(r)=1- {m(r)}/{r^2}$. Then the coupled static EYMCS equations of motion reduce to
\begin{eqnarray}
\label{eqs}
&&m'=\frac{1}{2}\alpha^2 
\left(
3r \bigg(
N w'^2 
+\frac{(w^2 -1)^2}{r^2}
\bigg)
+\frac{r^3}{\sigma^2}  V'^2 
\right),
~~\frac{\sigma'}{\sigma}=\frac{3\alpha^2 w'^2 }{2r},
\\
\nonumber
&&(r\sigma Nw')'=
\frac{2 \sigma  w(w^2 -1)}{r }
+8\kappa 
 V'(w^2 -1) ,
~~~\big (\frac{r^3 V'}{\sigma}\big )'= 24 \kappa (w^2 -1)w',
\end{eqnarray}
where
the prime indicates the derivative with respect to $r$ and $\alpha^2=16\pi G/(3g^2)$ 
(a factor of $1/g$
is also absorbed in $\kappa$).
The last equation above has the first integral
\begin{eqnarray}
\label{1SU2U1}
V'=\frac{\sigma}{r^3}\big(K+8 \kappa w(w^2-3)\big),  
\end{eqnarray}
with $K$ an integration constant.
 
The set of equations (\ref{eqs}) admit the scaling transformation 
\begin{eqnarray}
\label{ss1}
r\to \lambda r, ~~m\to \lambda^2 m,~~\sigma \to \sigma,~w\to w,~~V\to V/\lambda,
~~\kappa \to \lambda \kappa ~~{\rm and}~~\alpha\to \lambda \alpha~, 
 \end{eqnarray}
which is used in what follows to set $\alpha=1$.  
 
 We consider EYMCS black holes, with an event horizon located at $r=r_h>0$. 
 In the vicinity of the event horizon, one finds the following expansion of the solutions 
\begin{eqnarray}
\nonumber
&&m(r)=r_h^2+m_1(r-r_h)+\dots,
~
\sigma(r)=\sigma_h+\frac{3 \sigma_h w_1^2}{2 r_h}(r-r_h)+\dots,
\\
\label{exp-eh}
&&w(r)=w_h+w_1(r-r_h)+\dots,~
 V(r)=v_1(r-r_h)+\dots,
\end{eqnarray} 
where
$
v_1=\frac{\sigma_h(K+8\kappa (w_h^2-3))}{r_h^3},$
$
m_1=\frac{r_h^4 v_1^2 +3\sigma_h^2(1-w_h^2)^2}{2r_h \sigma_h^2}
$,
$
w_1= \frac{2(4\kappa r_h v_1+\sigma_h w_h)(1-w_h^2)}
{(m_1-2r_h)\sigma_h}.
$
while for large values of $r$, the expression of the solutions is
\begin{eqnarray}
\label{exp-inf}
 m(r)=M-\frac{Q^2}{r^2} +\dots,
~~
\sigma(r)=1-\frac{ J^2}{r^6}+\dots,
~~ w(r)=\pm 1+\frac{ J}{r^2}+\dots,~~V(r)=V_0-\frac{Q}{r^2}+\dots~.
\end{eqnarray} 
In the above relations, $\sigma_h,~w_h$ and $J,~M,~V_0$ are parameters given by numerics which fix all higher order terms, 
while $Q=K/2 \mp 8 \kappa$.

The only conserved quantities associated
with these solutions are the mass ${\cal M}=\frac{3\pi}{8 G}M$ and the
electric  charge ${\cal Q}=\frac{4 \pi^2 }{g}Q$, which is associated  with the $U(1)$ gauge symmetry generated
by $\Sigma_{56}$.  
Other quantities of interest are the chemical potential $\Phi=\frac{V_0}{g}$,
the Hawking temperature $T_H= \frac{1}{4 \pi} \sigma(r_h) N'(r_h)$ 
and the entropy $S=\frac{\pi^2 r_h^3}{2G}$.
 

\noindent{\textbf{~~~The solutions.--~}}
The moduli space of black hole solutions includes the Reissner-Nordstr\"om black hole (hereafter RN),  
described by $m(r)=M-\frac{Q^2}{r^2}$, $V(r)=V_0-Q/r^2$, $\sigma(r)=1$, $w(r)=\pm 1$. 
This  solution has an event horizon at $r_h= \bigg({\frac{M}{2}+\sqrt{\frac{M^2}{4}-Q^2}}\bigg)^{1/2}$,
which becomes extremal for $Q= \frac{M}{2}$.
We have found that for $\kappa \geq 1/8$ and a given $Q>0$,
the RN black hole presents an instability with respect to static non-Abelian perturbations,
for a critical value of the mass. 
This instability is found within the Ansatz (\ref{YMansatz}),
for values of the magnetic gauge potential
$w(r)$ close to the value $-1$ everywhere, $w(r)=-1+ \epsilon W(r)$.
The perturbation $W(r)$  starts from some nonzero value at the horizon and vanishes at infinity,
being a solution of the linear equation
\begin{eqnarray}
\label{stab1}
 r(r N W')'-4(1-\frac{8\kappa Q}{r^2})W=0~,
\end{eqnarray}
where $N=1-\frac{Q^2+r_h^4}{r_h^2r^2}+\frac{Q^2}{r^4}$.
The second term in (\ref{stab1}) shows the existence of an effective mass term $\mu^2$ for $W$ near the horizon, with 
$\mu^2\sim 1-8\kappa Q/r_h^2$. All physical solutions have $\mu^2<0$,  with $Q/r_h^2=U(\kappa)$ being a monotonic
function of the CS coupling constant $\kappa$. Then, for given $\kappa\geq 1/8$, an instability occurs for a critical
value of the mass to charge ratio of the RN solution, with $M/Q=(1+U^2)/U$.
As $\kappa\to 1/8$, one finds $U=1$, while $U\simeq 1/4\kappa$ for large $\kappa$. No
solutions of (\ref{stab1}) are found for $\kappa<1/8$, or for perturbations of the form $w(r)=+1+ \epsilon W(r)$,
 in which cases the effective mass for $W$ is always real. 

 \begin{figure}[t]
\begin{center}
\epsfysize=6.5cm
\mbox{\epsffile{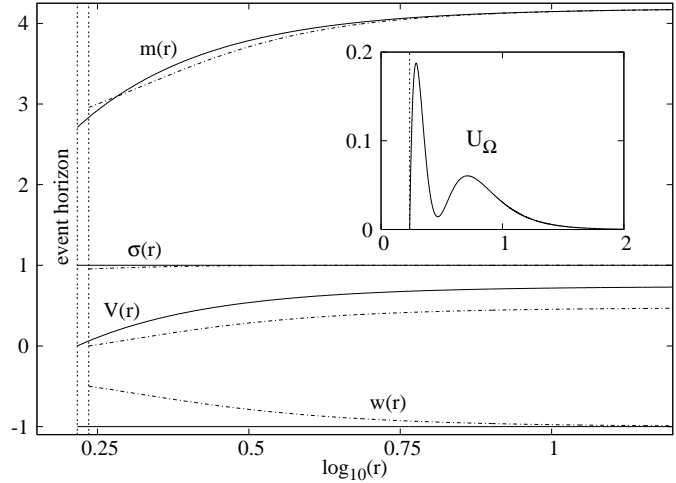}}
\caption{
The profiles of a typical non-Abelian (dotted line) solution with $\kappa=0.5$
is shown together with the corresponding Reissner-Nordstr\"om (RN)
black hole (solid line) with the same mass and electric charge parameters $M=4.186$ and $Q=2$,
respectively 
(the horizon radius of RN is smaller than that of the non-Abelian black hole). 
The intlet shows the corresponding potential for the perturbation equation
(\ref{schr}) below.  
}
\end{center}
\end{figure}

 This unstability signals the presence of a symmetry breaking branch of non-Abelian solutions bifurcating from
the RN black hole.  These nonperturbative configurations are found by solving numerically the Eqs. (\ref{eqs}), 
using a shooting method. In this work we have restricted
attention to solutions with a monotonic behaviour\footnote{Although there exist solutions where 
$w$ has local extrema, it is likely that these are always thermodynamically disfavoured 
because spatial oscillations in $w$ increase the total mass.}
of the magnetic gauge potential $w(r)$.
In contrast to other asymptotically flat non-Abelian black holes  \cite{89}, some of the  EYMCS 
solutions have no nodes in the magnetic gauge function $w(r)$.
A typical nodeless profile is shown in Fig. 1, together with the corresponding Abelian solution with the same mass and 
electric
charge.

 \begin{figure}[t]
\begin{center}
\epsfysize=6.5cm
\mbox{\epsffile{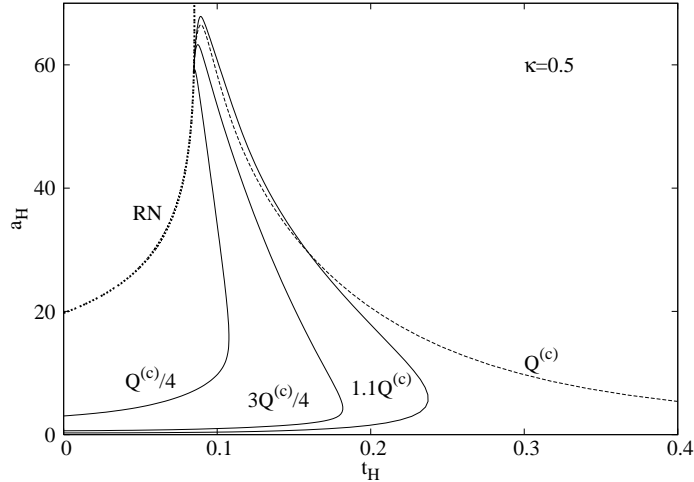}}
\caption{
The dimensionless horizon area  is plotted $vs.$ the scaled temperature   
for the non-Abelian  solutions with several values of the electric
charge and a given value of $\kappa$. The branch of Reissner-Nordstr\"om solutions is also shown. 
}
\end{center} 
\end{figure}

Solutions smoothly interpolating between the asymptotics (\ref{exp-eh}) and (\ref{exp-inf}) 
appear  to exist for any value of the CS coupling constant $\kappa \geq 1/8$.
In a canonical ensemble, the  non-Abelian black holes with a given $\kappa$
exist for a finite interval of  $r_h$ ($i.e.$ of the entropy) only. 
The detailed picture depends however on the ratio $Q/\kappa$,
with a critical value $Q^{(c)}=16\kappa$.  
For $Q \neq Q^{(c)}$,
the temperature reaches its maximum at some intermediate value of the event horizon radius.
Then a plot of the horizon area as a function of the temperature 
reveals  
the existence of several branches of non-Abelian solutions. 
The upper branch ends in the critical RN solution with  $r_h=\sqrt{Q/U(\kappa)}$.
The lower branch possesses always a positive specific heat, the Hawking temperature vanishing there 
for a minimal value  $r_h^{(min)}$  of the event horizon radius.
As $r\to r_h^{(min)}$,
 an extremal non-Abelian black hole solution with a regular horizon
is approached.  
For $Q <Q^{(c)}$ the near horizon expansion of the solutions implies
$r_h^{(min)}= {4\sqrt{3}  \kappa (1-w_h^2)}/{\sqrt{64 \kappa^2- w_h^2}},$
where $w_h$ satisfies the equation $(64 \kappa^2-w_h^2)Q^2+2\kappa(1+w_h)^2(128 \kappa^2(w_h-2)+w_h(w_h^2-2w_h+3))=0$;
for $Q >Q^{(c)}$, the limiting solution has $r_h^{(min)}=\sqrt{Q-16\kappa}$ and  $w_h=1$.
 Some of these features are shown in Fig. 2 where we plot the reduced area of the horizon $a_H=2\pi^2 r_h^3/Q^{3/2}$ 
as a function of the   dimensionless temperature 
$t_H$ for a fixed  CS coupling constant and several values of the charge parameter $Q$
 (both $a_H$ and $t_H$ are invariant under the  scaling symmetry (\ref{ss1})). 
The branch of RN solutions is also shown there.

\begin{figure}[t]
\begin{center}
\epsfysize=6.5cm
\mbox{\epsffile{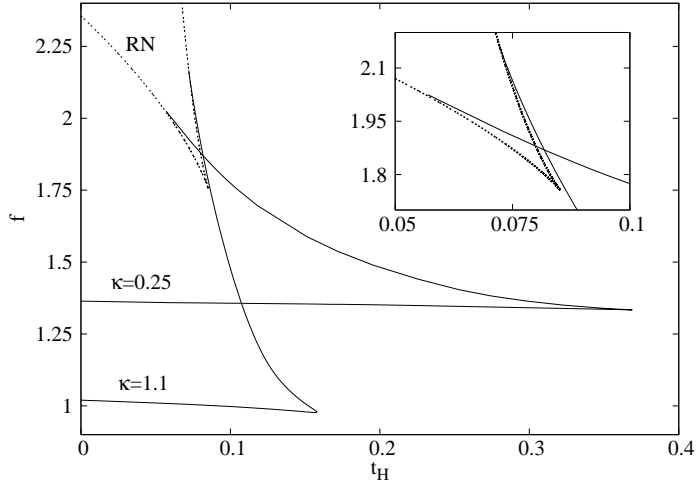}}
\caption{
The scaled free energy is plotted $vs.$ the scaled temperature for the Reissner-Nordstr\"om solutions (dotted
line) and the two non-Abelian sets of solutions
with different values of the Chern-Simons coupling constant $\kappa$ and the same charge parameter $Q=3.8$.}
\end{center}
\end{figure}

Furthermore, it turns out that
the free energy $F={\cal M}-T_H S$  
of a RN solution  is larger than the free energy
of a lower branch non-Abelian solution with the same temperature and electric charge, except for 
 configurations with $\kappa$ close to $1/8$ and small enough values of the charge, $Q \lesssim Q^{(c)}/3$.
Therefore the non-Abelian black holes are generically preferred\footnote{Note that 
the
non-Abelian solutions with large enough temperature have
no RN counterparts, see Fig. 3.}.
This reveals the existence of a second order phase transition between the RN solutions and the 
non-Abelian solutions. 
These aspects are exhibited in Fig. 3 where the dimensionless free energy $f=G F/Q$  is 
plotted as a function of the  dimensionless temperature $t_H= T_H \sqrt{  Q} $ for two values of $\kappa$. 
Moreover, as seen $e.g.$ in Fig. 1, for the same values of the mass and electric charge, 
the RN solution typically  has a smaller event horizon radius (and thus a smaller entropy),
than the non-Abelian black hole.
The parameter $J$ which enters the large $r$ asymptotics of the magnetic gauge potential $w(r)$
increases from zero (for the critical RN solution) to a maximal value approached at $T_H=0$. 

The overall picture is somehow different for  $Q=Q^{(c)}$, in which case,
despite the presence of an electric charge, the  non-Abelian black holes behave in a 
similar way to the vacuum Schwarzschild-Tangherlini 
solution, with a single branch of thermally unstable configurations, see Fig. 2. 
In the limit $r_h\to 0$, these black holes 
approach a set of globally regular particle-like solutions, whose mass is an almost linear function of $\kappa$.


In the special case $\kappa=1/8$, following the approach in \cite{Gibbons:1993xt},
one finds the following exact solution of the EYMCS equations within the general Ansatz (\ref{metric-gen}), (\ref{YMansatz}):
\begin{eqnarray}
\label{ex-sol1}
w=\frac{J-2 r^2}{J+2 r^2},~~
V=  F , ~~
f_0=F^2 ,~~
f_1=\frac{f_2}{r^2}=\frac{1}{F},~~{\rm with}~~
F=\left(1+\frac{Q-2}{r^2}+\frac{2(r^2+J)}{(r^2+J/2)^2}\right)^{-1},~
\end{eqnarray}
where $Q\geq 2$ and $J$ are arbitrary parameters.
This describes an extremal black hole with non-Abelian hair, the regular event horizon being at $r=0$
 (in these coordinates).
The mass, electric charge and entropy of this non-Abelian deformation of the extremal RN solution are 
${\cal M}={\cal Q}/g=\frac{4\pi^2}{g^2}Q$,
$S=8\pi^3 (Q-2)^{3/2}/ 3g^2 $ 
(note that $J$  does not enter any physical quantity). 
In the limit
$Q\to 2$, the solution (\ref{ex-sol1}) describes a particle-like soliton with a regular origin.

\noindent{\textbf{~~~The stability of solutions.--~}
 The fact that we have found nodeless solutions suggest the existence of non-Abelian configurations
 stable against spherically symmetric perturbations. 
In examining such time-dependent fluctuations, we consider the metric Ansatz (\ref{metric-gen}) with
$ f_1=1- \frac{m(r,t)} {r^2},~f_2=r^2,~f_0=(1- \frac{m(r,t)} {r^2})\sigma^2(r,t)$, 
the YM Ansatz (\ref{YMansatz}) and 
the following perturbed
variables\footnote{The corresponding problem for the SU(2)
EYM system has been considered in \cite{Okuyama:2002mh}.}   
\begin{eqnarray}
\nonumber
&& m(r,t)=m(r)+\epsilon m_1(r)e^{i \Omega t},~ 
\sigma(r,t)=\sigma(r)(1+\epsilon \sigma_1(r)e^{i \Omega t}),
\\
\nonumber
&& w(r,t)=w(r)+\epsilon w_1(r)e^{i \Omega t},~ 
V(r,t)=V(r)+\epsilon V_1(r)e^{i \Omega t}.
\end{eqnarray}
One finds that in the system
of linearised EYMCS equations,  the functions $m_1,V_1$ and $\sigma_1$
can be eliminated in favor of $w_1(r)$, leading to a single Schr\"odinger equation
for $w_1$:  
\begin{eqnarray}
\label{schr}
-\frac{d^2 \chi}{d \rho^2}+U_{\Omega}(\rho)\chi=\Omega^2 \chi,
\end{eqnarray}
where $\chi=w_1 \sqrt{r}$, $dr/d \rho=N \sigma $ 
and
\begin{eqnarray}
\nonumber
U_{\Omega} =\frac{N\sigma^2}{r^2}
\bigg[
6(w^2-w'^2-\frac{1}{6})
-\frac{5N}{4}
+12 (w^2-1)\frac{ww'}{r}
+\frac{(1-w^2)^2}{r^2}
(
\frac{9}{2}w'^2+
192 \kappa^2 -\frac{3}{4}
)
\\
\label{potential}
+\frac{16\kappa V'}{\sigma}(rw -3 (1-w^2)w')
-\frac{r^2 V'^2(1-6 w'^2)}{4\sigma^2}
\bigg].
\end{eqnarray}  
The potential above is regular in the entire range $-\infty<\rho<\infty$.
Near the event horizon, one finds $U_{\Omega}\to 0$; for large values of $\rho$ the potential
is positive and  bounded. Standard results from quantum mechanics \cite{Messiah} further imply that 
there are no negative eigenvalues for $\Omega^2$ (and then no unstable modes) 
if the potential $U_{\Omega}$ is everywhere positive.

Although the potential (\ref{potential}) is  not positive definite for all values of $Q,\kappa$,
we have found numerically that the condition  $U_{\Omega} > 0$ 
is fulfilled by some of the nodeless solutions  
(see   the  Fig. 1 for a such a configuration).  
Therefore at least some of our solutions
are linearly stable. 
The full picture is, however, quite complicated and a detailed discussion  will be
presented elsewhere. 
We note only that, by using the approach in \cite{Boschung:1994tc} we have found that the EYMCS solutions
with one node in $w(r)$ we have constructed are indeed unstable. 

\noindent{\textbf{~~~Further remarks.--~}} 
The main purpose of this work was to provide an example
of stable, non-Abelian black holes  without scalar fields, in asymptotically Minkowski spacetime. 

One should note that, despite the different asymptotic structure of spacetime and the different 
horizon topology, the  solutions in this work have some similarities with the
colorful black holes with charge in Anti-de Sitter (AdS) space \cite{Gubser:2008zu,Manvelyan:2008sv},
which provide a model of holographic superconductors.
In both cases, an Abelian gauge symmetry is spontaneously broken near a black hole horizon
with the appearance of a condensate of non-Abelian gauge fields  there,
leading to
a second order phase transition. 
It remains an interesting open problem to clarify if the asymptotically flat EYMCS black holes
may also provide useful analogies to phenomena observed in condensed matter physics.
However, given the presence of several branches, the picture is more 
complicated for asymptotically flat solutions 
and we could not find so far simple universal relations between the relevant parameters
as those found in \cite{Gubser:2008zu} in the AdS case. 

The black holes in this work admit a straightforward generalisation with a cosmological constant.
For $\Lambda<0$, the basic properties of asymptotically AdS configurations were discussed in \cite{Brihaye:2009cc}.
In this case the EYMCS model with a special value of $\kappa$ can be thought of as a truncation
of the ${\cal N}=8,~d=5$ gauged supergravity \cite{Gunaydin:1985cu, Cvetic:2000nc}, 
with all scalars there taking constant values.
We have found that most of the properties of the asymptotically flat solutions, in 
particular the existence of stable configurations, hold also in the AdS case.
In this case the solutions within the restricted $SO(4)\times O(2)$
Ansatz (\ref{YMansatz}) contain already all relevant features of the full $SO(6)$ configurations.
 
While the existence of hairy non-Abelian black holes with AdS asymptotics is not
a surprise in view of the results in  \cite{Okuyama:2002mh}, \cite{Winstanley:1999sn},
we have found that, 
the $\Lambda=0$ solutions 
 presented here admit
generalisations with de-Sitter asymptotics ($\Lambda>0$)  as well.
These black holes,  possesing a regular cosmological event horizon at $r=r_c>r_h$, were constructed
within the same
Ansatz as in the asymptotically flat case.
By using an  inflationary coordinate system, one finds also the following generalisation of 
(\ref {ex-sol1}), with   $\kappa=1/8$ again and 
\begin{eqnarray}
\label{exact-solution-dS}
ds^2=\frac{e^{2H t}}{F(r,t)} (dr^2+r^2d\Omega_3^2)- F^2(r,t)dt^2,~
w(r)=\frac{J-2 r^2}{J+2 r^2},~
V(r,t)= {F(r,t)},
\end{eqnarray}
where $F(r,t)^{-1}= 1+ e^{-2H t}\left(\frac{Q-2}{r^2}+\frac{2(r^2+J)}{(r^2+J/2)^2}\right) $
and $\Lambda=6H^2$.
This configuration describes a non-Abelian deformation of the  
extremal RN-de Sitter black holes in \cite{Kastor:1992nn}.
By using the methods in \cite{Kastor:1992nn}, \cite{Astefanesei:2003gw},
one finds that (\ref{exact-solution-dS}) shares the basic features of the $J=0$ Abelian configuration. 

Finally, we conjecture that similar EYMCS solutions
exist in higher (odd) dimensions, the features encountered here being universal.
\\
\\
\noindent{\textbf{~~Acknowledgements.--}} 
This work is carried out in the framework of Science Foundation Ireland (SFI) project
RFP07-330PHY. 
YB is grateful to the
Belgian FNRS for financial support.

\end{document}